\documentclass[%
reprint, twocolumn, superscriptaddress,
showpacs,preprintnumbers,
nofootinbib,
 amsmath,amssymb,
 aps,
pra,
]{revtex4}
\usepackage{mathrsfs}

\usepackage{graphicx}% Include figure files
\usepackage{dcolumn}% Align table columns on decimal point
\usepackage{bm}% bold math
\usepackage{color} % change color of text
\usepackage{CJK}
\def\la{{\langle}}
\def\ra{{\rangle}}

\newcommand{\beq}{\begin{equation}}
\newcommand{\eeq}{\end{equation}}
\newcommand{\beqa}{\begin{eqnarray}}
\newcommand{\eeqa}{\end{eqnarray}}

\begin{document}
\title{Fast shuttling of a trapped ion in the presence of noise}
\author{Xiao-Jing Lu}
\affiliation{Department of Physics, Shanghai University, 200444
Shanghai, People's Republic of China}
\affiliation{Departamento de Qu\'{\i}mica F\'{\i}sica, UPV/EHU, Apdo.
644, 48080 Bilbao, Spain}
\author{J. G. Muga}
\affiliation{Departamento de Qu\'{\i}mica F\'{\i}sica, UPV/EHU, Apdo.
644, 48080 Bilbao, Spain}
\affiliation{Department of Physics, Shanghai University, 200444
Shanghai, People's Republic of China}\author{Xi Chen}
%\email{xchen@shu.edu.cn}
\affiliation{Department of Physics, Shanghai University, 200444
Shanghai, People's Republic of China}
\author{U. G. Poschinger}
\affiliation{QUANTUM, Institut f\"ur Physik, Universit\"at Mainz, D-55128 Mainz, Germany}
\author{F. Schmidt-Kaler}
\affiliation{QUANTUM, Institut f\"ur Physik, Universit\"at Mainz, D-55128 Mainz, Germany}
\author{A. Ruschhaupt}
\affiliation{Department of Physics, University College Cork, Ireland}
%\author{..}
%\affiliation{..}
%
%\affiliation{$^{2}$Department of Physics, Shanghai University,
%200444 Shanghai, P. R. China}
%
%\affiliation{$^{1}$Departamento de Qu\'{\i}mica-F\'{\i}sica,
%UPV-EHU, Apdo 644, 48080 Bilbao, Spain}

%
\begin{abstract}
We theoretically investigate the motional excitation of a single ion
caused by spring-constant and position  fluctuations of a
harmonic trap during trap shuttling processes. A detailed study of the
sensitivity on noise for several transport protocols and noise spectra
is provided.
The effect of slow
%systematic
spring-constant drifts
%deviations
is also analyzed.
Trap trajectories that minimize the excitation are designed combining  invariant-based inverse engineering,
perturbation theory, and optimal control.
\end{abstract}
\pacs{37.10.Ty, 03.67.Lx}
%\pacs{37.10.Gh, %Atom traps and guides
%02.30.Yy, % Control theory
%03.65.Ca, %Formalism
%03.65.Nk % Scattering theory
%37.10.Ty Ion trapping
%}
\maketitle
%
%
%
%
%\tableofcontents
%
\section{Introduction}
%
%Fast and stable ion transport is the key step in the quantum
%computer.
A quantum information processing architecture based on shuttling
individual or small groups of ions among different storing or
processing sites requires fast transporting techniques that avoid
decoherence and excitations at the arrival zone
\cite{Wineland2002,Rowe,Wineland2006}.
A promising research and technological avenue \cite{Roos,Monroe} has been opened by recent experiments \cite{Rowe,Home,transna}
that demonstrate  the feasibility  of a transport-based architecture, even beyond (faster than) the adiabatic regime \cite{Bowler,Schmidt}.
Such experiments on transport and fast splitting of ion crystals have been performed with optimized time-dependent control voltages and the outcome is analysed with spectroscopy precise at the level of single motional quanta \cite{Bowler,Schmidt}.
A fundamental limit to the shuttling speed which can be achieved at a desired final excitation is given by the unavoidable presence of noise.
Electric field noise in Paul traps has been characterized experimentally in Ref. \cite{TURCHETTE2000} by monitoring the heating out of the motional ground state. It was found that the corresponding noise level exceeds the limit given by Johnson noise by several orders of magnitude, an effect that has been termed
\textit{anomalous heating}. In a resting trap, it has been shown \cite{Lamoreaux} that the heating rate is determined by the noise power spectral density at the trap
frequency. This does not necessarily hold for shuttling operations, where a broader part of the noise spectrum
%. The corresponding protocols rely on well-defined trap parameters. Thus, both noise
%at frequencies below the trap frequency
and slow drifts of the trap parameters can compromise the shuttling result producing undesired
excitation.
%in terms of residual energy.
\\
On many ion trap experiments, the frequency dependence of the electric field noise spectral density  $S(\Omega)$
has been investigated by measuring heating rates for varying trap frequencies, and commonly a polynomial scaling $S(\Omega)  \propto \Omega^{-\alpha}$ is observed. While a wide range of exponents $\alpha$ between -1 and 6 have been reported \cite{Brownnutt}, in many cases a behavior consistent with flicker noise $\alpha\approx 1$ is observed. This indicates that a variety of noise spectra can occur and that resonances of technical origin can play a role. Fast shuttling operations ultimately require rapidly changing voltage waveforms \cite{ BAIG2013, BOWLER2013}, which strongly restricts the possibility to mitigate noise by filtering. This leads us to the conclusion that it is worthwhile to investigate the sensitivity of shuttling protocols for colored noise. We also consider drifts of trap parameters, which are slow on the timescales of the trap period and the durations of shuttling operations. These drifts can be characterized by monitoring the trap frequency over time. On a trap
similar to the design used in \cite{Schmidt}, we find long-time variations of the trap frequency of up to 5\%.
These variations can be caused by drift of the trap voltages, thermal expansion of the trap and charging of the trap itself.\\
Faster than adiabatic trap trajectories without final excitation may be designed using invariant-based
inverse engineering \cite{Erik1,Erikcond,transOCT,2ions,Mainz2}. This technique produces families of trajectories
as reviewed in Sec. \ref{II}.
It is possible to choose among them the ones that optimize some variable of interest, for example
the time average of the transient energy or of the displacement between the trap center and the center
of mass \cite{transOCT}.
In the field of internal state control this type of multiplicity has been used to design protocols
that minimize the effects of noise \cite{noise1,noise2}, and we shall apply this idea to ion transport too.
In Sec. \ref{noise}, we shall consider two basic
types of noise that affect a moving harmonic trap: spring-constant
fluctuations and position fluctuations around the ideal trajectory
(trap shaking). A basic challenge in transport-protocol design is to mitigate
or suppress noise or systematic errors and their  effects
on the final state fidelities. Our aim is to characterize
and minimize noise effects by finding optimal transport trajectories
and strategies. \\
%The excitation of position noise turns out to be independent of the trajectory
%followed by the trap so there is nothing that can be done against it other than
%speeding up the transport as much as possible.
We provide general results for the final excitation energy for different noise power spectra by using a perturbative master-equation approach.
A detailed study of the three relevant cases of white noise (flat spectrum), brown noise (Ornstein-Ulhenbeck process
with Lorentzian spectrum) and pink or flicker noise (1/frequency spectrum in a frequency range) is performed.

In Sec. \ref{error}, trajectories are found that
minimize the effect of a systematic (constant, not random) spring
constant error, and, finally,  we discuss how our theoretical results may be implemented experimentally.
\section{Invariant-based inverse engineering method\label{II}}
The harmonic transport of one ion
is described here by the effective 1D Hamiltonian
\beq
\label{Hamiltonian}
H_0 (t)= \frac{\hat{p}^2}{2 m} +
\frac{1}{2}m \omega^2 [\hat{q}-q_0(t)]^2,
\eeq
where $\hat{q}$ and $\hat{p}$ are the position and momentum
operators, $\omega/(2\pi)$ is the  frequency of the
trap, and $q_0(t)$ its center. The corresponding quadratic-in-momentum
Lewis-Riesenfeld invariant \cite{LR,LL,DL} is given in this case (up
to an arbitrary multiplicative constant) by \cite{Erik1}
\beq
\label{inva}
I (t) = \frac{1}{2m}(\hat{p}-m\dot{q}_c)^2 +\frac{1}{2}m \omega^2 [ \hat{q}-q_c (t)]^2,
\eeq
where the function $q_c (t)$ must satisfy the auxiliary equation
\beq
\label{classical}
\ddot{q}_c+\omega^2(q_c-q_0)=0
\eeq
to guarantee the invariant condition
\beq
\frac{d I(t)}{d t} \equiv \frac{\partial I(t)}{ \partial t} +\frac{1}{i \hbar} [I(t), H_0(t)] =0.
\eeq
The expectation value of $I(t)$ remains constant
for solutions of the time-dependent Schr\"odinger equation $i \hbar
\partial_t\Psi (q,t) = H_0(t) \Psi (q,t)$. They can be expressed in terms of independent
``transport modes'' $\Psi_n(q,t)=e^{i\alpha_n} \psi_n(q,t)$ as
%
%\beqa
$
\Psi(q,t) = \sum_n c_n e^{i\alpha_n} \psi_n(q,t),
$
%\eeqa
%
where $n=0,1,...$; $c_n$ are time-independent coefficients;  and $\psi_n(q,t)$ are the
orthonormal eigenvectors of the invariant $I(t)$ satisfying $I(t)\psi_n(q,t)= \lambda_n\psi_n(q,t)$,
with real time-independent $\lambda_n$.
The Lewis-Riesenfeld phase is
\beq
\label{LRphase}
\alpha_n (t) = \frac{1}{\hbar} \int_0^t \Big\langle \psi_n (t') \Big|
i \hbar \frac{\partial }{ \partial t'} - H_0(t') \Big| \psi_n (t')  \Big\rangle d t'.
\eeq
For the harmonic trap considered here \cite{Erik1},
%
%\beqa \label{psin}
%\psi_n(q,t) &=& \frac{1}{(2^n n!)^{1/2}} \left( \frac{m \omega}{\pi \hbar}\right)^{1/4}  \exp{\left[- \frac{m \omega}{2\hbar} (q-q_c)^2\right]} \nonumber \\
% &\times& \exp{\left(i \frac{ m\dot{q}_c q}{\hbar}\right)}  H_n \left[\left(\frac{m \omega}{\hbar}\right)^{1/2}(q-q_c)  \right],
%\eeqa
%
%%
\beq
\label{psin} \psi_n(q,t) = \exp{\left(i \frac{ m\dot{q}_c
q}{\hbar}\right)}\psi_n^0(q-q_c),
\eeq
where $\psi_n^0(q)$ are the eigenstates of Eq. (\ref{Hamiltonian}) for $q_0(t)=0$.
%, $H_n$ is a Hermite polynomial.
%where $H_n$ is a Hermite polynomial.
Note that $q_c$ is the center
of mass of the transport modes obeying the classical Newton equation
(\ref{classical}). %%
%\beqa \label{psin}
%\psi_n(q,t) &=& \frac{1}{(2^n n!)^{1/2}} \left( \frac{m \omega}{\pi \hbar}\right)^{1/4}  \exp{\left[- \frac{m \omega}{2\hbar} (q-q_c)^2\right]} \nonumber \\
% &\times& \exp{\left(i \frac{ m\dot{q}_c q}{\hbar}\right)}  H_n \left[\left(\frac{m \omega}{\hbar}\right)^{1/2}(q-q_c)  \right],
%\eeqa
%%

Suppose that the harmonic trap is displaced from $q_0 (0) =0$ to
$q_0 (T) =d$ in a shuttling time $T$. The trajectory $q_0 (t)$ of the trap can
be inverse engineered by designing first an appropriate classical
trajectory $q_c (t)$. To guarantee the commutativity  of $I(t)$ and
$H_0(t)$ at $t=0$ and $t=T$, which implies the mapping between initial and final
trap eigenstates without final excitation,
we set the conditions \cite{Erik1}
\beqa
\label{conq}
q_0(0)=q_c(0)=0,~~ \dot{q}_c(0)=0,
\nonumber
\\
 q_0(T)=q_c(T)=d,~~\dot{q}_c(T)=0.
\eeqa
The additional conditions
\beq \label{conqdd}
\ddot{q}_c(0)=0,~~\ddot{q}_c(T)=0
\eeq
may be imposed to avoid sudden jumps in the trap position. However
discontinuities of $\ddot{q}_c(t)$ may in general be allowed: they
correspond to ideal instantaneous trap displacements inducing a
sudden finite jump of the acceleration, whereas the velocity
$\dot{q}_c$ and the trajectory $q_c$ remain continuous. In the
following, we  consider for simplicity the transport of the single
mode $n$ in the noiseless limit ($n=0$ in the numerical examples),
and examine the excitation energy of the system energy due to noise
or errors, as well as ways to suppress or minimize it.

%%%%%%%%%%%%%%%%%%%%%%%%%%%%%%%%%%%%%%%%%%%%%%%%%%%%%%%%%%%%%%%%%%%%%%%%%%%%%%%%%%
%
%
%
%
%
\section{Noise}\label{noise}
To study the effect of the noise we follow the master equation treatment in
\cite{diosi1,diosi2,strunzpla}. The Hamiltonian is assumed to be
of the form
\beq
H (t)=\frac{\hat{p}^2}{2 m} + \frac{1}{2}m \omega^2
[\hat{q}-q_0(t)]^2+Lx(t),
\label{hami}
\eeq
where $L$ is a system operator coupling to the environment. The
fluctuating variable $x(t)$ satisfies
\beq
{\cal{E}}[x(t)]=0,~~{\cal{E}}[x(t)x(s)]=\alpha(t-s),
\eeq
where $\alpha(t-s)$ is the correlation function of the noise
and $\cal{E}[...]$ the statistical expectation.
%The
%power spectrum of the noise is given by
%
%\beq S(\Omega)={\cal{E}}[X(\Omega)X^\ast(\Omega)],
%\eeq
%
%where $$X(\Omega)=\int_{-\infty}^{\infty}e^{i\Omega t}x(t)dt.$$
The
correlation function and the spectral power density are related by
the Wiener-Khinchin theorem,
\beqa
S(\Omega)&=&\frac{1}{2\pi}\int_{-\infty}^\infty
\alpha(\tau)\cos(\Omega\tau) d \tau,
\label{som}
\\
%
%and
%
\alpha(\tau)&=&\int_{-\infty}^\infty S(\Omega)\cos(\Omega\tau) d
\Omega.
\eeqa
By expanding in the ratio between environmental correlation time and the typical time scale of the system
\cite{yuting}, a closed  master equation can
be derived retaining first order corrections to the Markovian limit,
\beq \label{ME}
\frac{d}{dt}\rho=-\frac{i}{\hbar}[H_0,\rho]+\frac{1}{\hbar}[L,\rho\bar{O}(t)^{\dag}]
-\frac{1}{\hbar}[L^{\dag},\bar{O}(t)\rho],
\eeq
where
\beq
\bar{O}(t)=\frac{1}{\hbar}g_0(t)L-\frac{i}{\hbar^2}g_1(t)[H_0,L]-\frac{g_2(t)}{\hbar^3}[L^{\dag},L]L,
\eeq
and
\beqa g_0(t)&=&\int_0^t\alpha(t-s)d s,
\\
g_1(t)&=&\int_0^t\alpha(t-s)(t-s)d s,
\\
g_2(t)&=&\int_0^t\int_0^s\alpha(t-s)\alpha(s-u)(t-s)du\,ds.
\eeqa
%%
%\beqa g_0(t)&=&\int_0^\infty \frac{S(f)}{2\pi f}\sin2\pi ft d f,
%\\
%g_1(t)&=&\int_0^\infty\frac{S(f)}{4\pi^2 f^2}(2\pi ft\sin2\pi
%ft+\cos2\pi ft-1)d f,
%\\
%g_2(t)&=&\int_0^\infty \frac{S(f)}{8\pi^3 f^3}(\pi^2 f^2t^2\sin2\pi
%ft+\frac{\pi ft}{2}\cos2\pi ft-\frac{1}{4}\sin2\pi ft)d f. \eeqa
%%
We insist that the master equation
(\ref{ME}) is valid on the condition that the noise correlation time
is small compared to the typical system time scales, so that $g_1$ and $g_2$
terms must be corrections to the dominant $g_0$ term.
%which also means the color noise is not far away from the white noise.
%
%
%
%
\subsection{Spring constant noise}
We consider now a fluctuating spring constant by setting
$L=\frac{1}{2}m\omega^2(\hat{q}-q_0)^2$. Then
\beq
\bar{O}(t)=\frac{m\omega^2}{2\hbar}(t)g_0(t)(\hat{q}-q_0)^2+\frac{2}{
m\hbar}g_1(t)\!\!\left[\hat{p}q_0-\frac{1}{2}(\hat{p}\hat{q}+\hat{q}\hat{p})\right]\!,
\eeq
and the master equation (\ref{ME}) becomes
\beqa
\label{f-masterEq}
\frac{d}{dt}\rho=&-&\frac{i}{\hbar}[H_0,\rho]-\frac{m^2\omega^4}{4\hbar^2}g_0(t)\left[(\hat{q}-q_0)^2,[(\hat{q}-q_0)^2,\rho]\right]
\nonumber\\&-&\frac{m\omega^4}{2\hbar^2}g_1(t)\!\!\left[(\hat{q}-q_0)^2,[\hat{p}q_0-\frac{1}{2}(\hat{p}q+q\hat{p}),\rho]\right]\!.
\eeqa
Using time-dependent perturbation theory for the master equation
we may write the density operator as
(for an alternative non-perturbative approach see the Appendix A)
\beqa
\rho(T)&\simeq&\rho_0(T)+\frac{m^2\omega^4}{4\hbar^2}\int_0^Tg_0(t)\widetilde{U}_0(T,t)\widetilde{J}_1(t)\rho_0(t)dt
\nonumber
\\
&+&\frac{m\omega^4}{2\hbar^2}\int_0^Tg_1(t)\widetilde{U}_0(T,t)\widetilde{J}_2(t)\rho_0(t)dt,
\eeqa
where the subscript $``0"$ represents noiseless unitary dynamics,
$\rho_0(T)=|\Psi_n(T)\rangle\langle\Psi_n(T)|$, and
$\widetilde{U}_0(T,t)$ is the noiseless evolution superoperator,
i.e.,
\beq
\rho_0(t)=\widetilde{U}_0(t,t')\rho_0(t')=U_0(t,t')\rho_0(t')U_0^{\dag}(t,t'),
\eeq
where $U_0(t,t')$ is the noiseless evolution operator. $\widetilde{J}_1(t)$
and $\widetilde{J}_2(t)$ are superoperators,
\beqa
\!\!\widetilde{J}_1(t)\rho_0(t)\!&\!=\!&\!-[(\hat{q}-q_0)^2\!,\![(\hat{q}-q_0)^2,\rho_0(t)]],
\\
%
%and
%
%\beq
\!\!\widetilde{J}_2(t)\rho_0(t)&=&\!-\!\bigg[\!(\hat{q}-q_0)^2\!,\!\bigg[\hat{p}q_0\!-\!
{{\frac{(\hat{p}\hat{q}+\hat{q}\hat{p})}{2}}},\rho_0(t)\!\bigg]\!\bigg]\!.
\eeqa
A detailed calculation gives the final energy corresponding in the noiseless limit to the $n_{th}$ mode,
\beqa
\label{f-energy}\nonumber \la H_0(T)\ra_n &=&
tr[H_0(T)\rho(T)] \simeq \la\Psi_n(T)|H_0(T)|\Psi_n(T)\ra
\\
\nonumber
&+&\frac{m^2\omega^4}{4\hbar^2}\int_0^Tg_0(t)\la\Psi_n(t)|\widetilde{J}_1(t)H'(t)|\Psi_n(t)\ra
dt
\\\nonumber
&+&\frac{m\omega^4}{2\hbar^2}\int_0^Tg_1(t)\la\Psi_n(t)|\widetilde{J}_3(t)H'(t)|\Psi_n(t)\ra
dt
\\\nonumber
&=&E_n+\hbar\omega^3\left(n+\frac{1}{2}\right)\int_0^Tg_0(t)dt,
\\\nonumber
&+&m\int_0^T[g_0(t)\ddot{q}_c^2(t)+\omega^2g_1(t)\dot{q}_c(t)\ddot{q}_c(t)]dt,
\eeqa
where $E_n=(n+1/2)\hbar\omega$,
$$\widetilde{J}_3(t)H'(t)=-\left[\hat{p}q_0-\frac{1}{2}(\hat{p}\hat{q}+\hat{q}\hat{p}),[(\hat{q}-q_0)^2,H'(t)]\right],
$$
and $H'(t)=U_0^{\dag}(T,t)H_0(T)U_0(T,t)$.

The following subsections deal with different noise types according to their spectrum.
We pay much attention to white noise because our method is perturbative, so understanding this
reference case in depth is fundamental. In addition, white noise is amenable of analytical treatment
and explicit optimization of trap trajectories.
%
%%%%%%%%%%%%%%%%%%%%%%%%%%%%%%%%%%%%%%%%%%%%%%%%%%%%%%%%%%%%%%%%%%%%%%%%%%%%%%%%%%%%%%%%%%
%
%
%
\subsubsection{White noise}
The correlation function for white noise is
$\alpha(\tau)=\gamma\delta(\tau)$, and the corresponding power
spectrum is constant, $S(\Omega)=\frac{\gamma}{2\pi}$. Here $\gamma$
scales the noise and
\beqa
g_0(t)=\gamma/2,~g_1(t)=0.
\eeqa
The instantaneous energy in Eq. (\ref{f-energy}) can be expressed as
\beq
\la H_0(T)\ra_n=E_n+\gamma
G(T),
\eeq
where
\beqa
G(T)=\frac{m}{2}\int_0^T\ddot{q}_c^2(t)dt+\frac{\hbar
\omega^3}{2}\left(n+\frac{1}{2}\right)T.
\eeqa

The excitation energy is $E_e=\gamma G(T)$.
The first term of $G(t)$ contains an integral of
$\ddot{q}_c(t)$ and the mass of the ion, it reflects the fact that
larger displacements from the trap center, see Eq. (\ref{classical}),
increase the effect of spring constant fluctuations.
The second term depends on
trap frequency and the final time, and it is
independent of the trajectory, so it can only be reduced by speeding
up the transport.
For fixed $T$, however, it is possible to design the trajectory
$q_c(t)$ to make $G(T)$ as small as possible and minimize the
integral. We shall now consider four different protocols. Examples
of the corresponding trap trajectories $q_0(t)$ are provided in Fig. 1.

%%%%%%%%%%%%%%%%%%%%%

\begin{figure}[h]
\begin{center}
\scalebox{0.6}[0.6]{\includegraphics{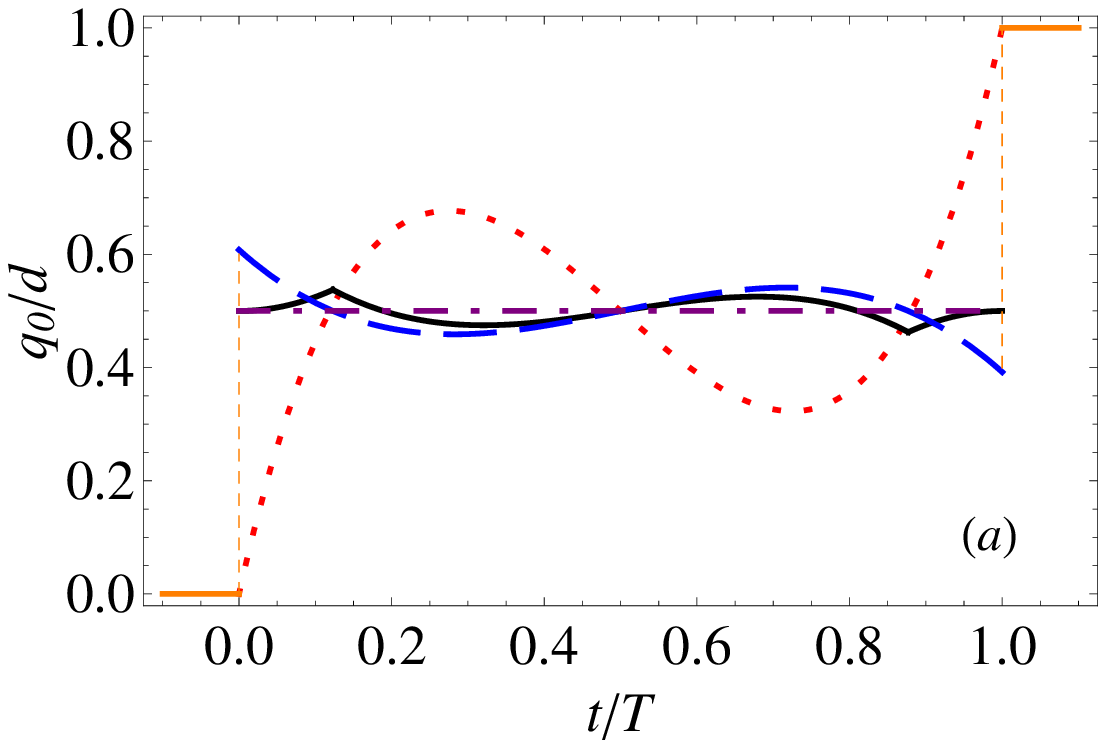}}
\scalebox{0.6}[0.6]{\includegraphics{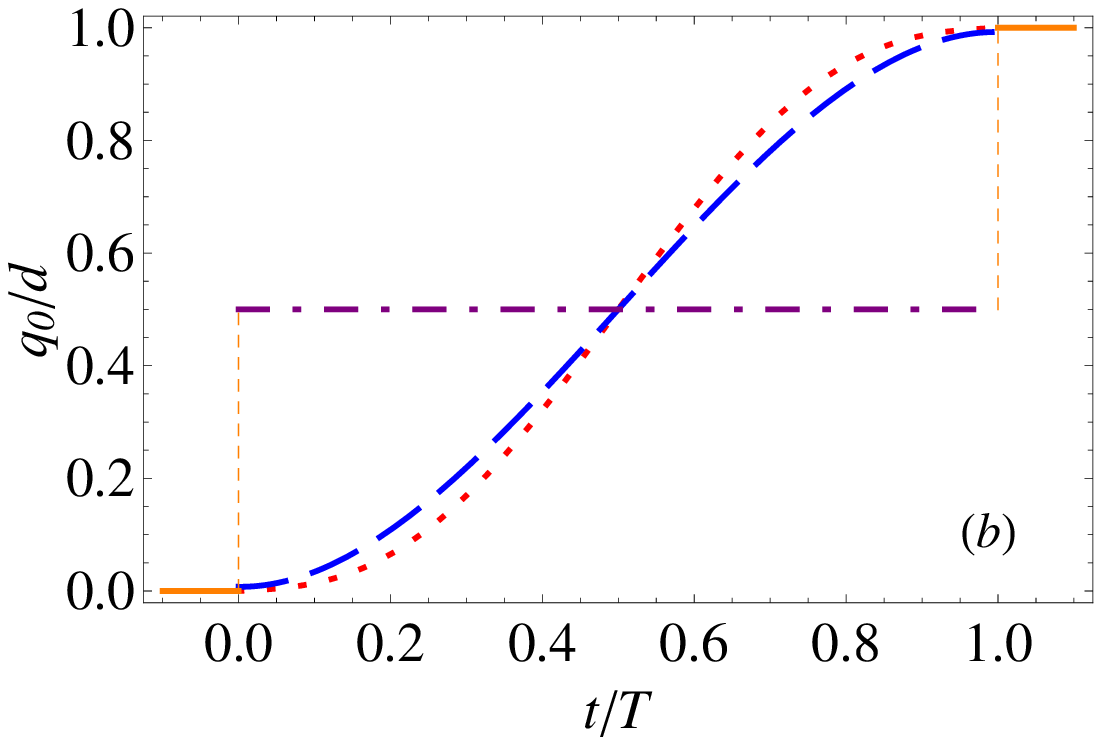}} \caption{(Color
online) Comparison of trap trajectories $q_0$ between the initial
and final trap positions (yellow solid segments). Polynominal
protocol (red dotted line); bounded optimal (black solid line, only
in (a)); unbounded optimal (blue dashed line, the jump at the
boundary times is $6d/(\omega^2T^2)$); bang-bang  (dot-dashed purple
line).  In (a) $T=T_0/2$ ($T_0$ is the oscillation period); in (b)
$T=9 T_0/2$. $\delta=0.5$ $d$, mass of $^{40}$Ca$^+$,
%m=6.64 \times10^{-26}$ kg,
initial state in $n=0$, $\omega = 2 \pi \times 1.4$ MHz, $d=280$
$\mu$m.} \label{figq0}
\end{center}
\end{figure}

\begin{figure}[h]
\begin{center}
\scalebox{0.65}[0.65]{\includegraphics{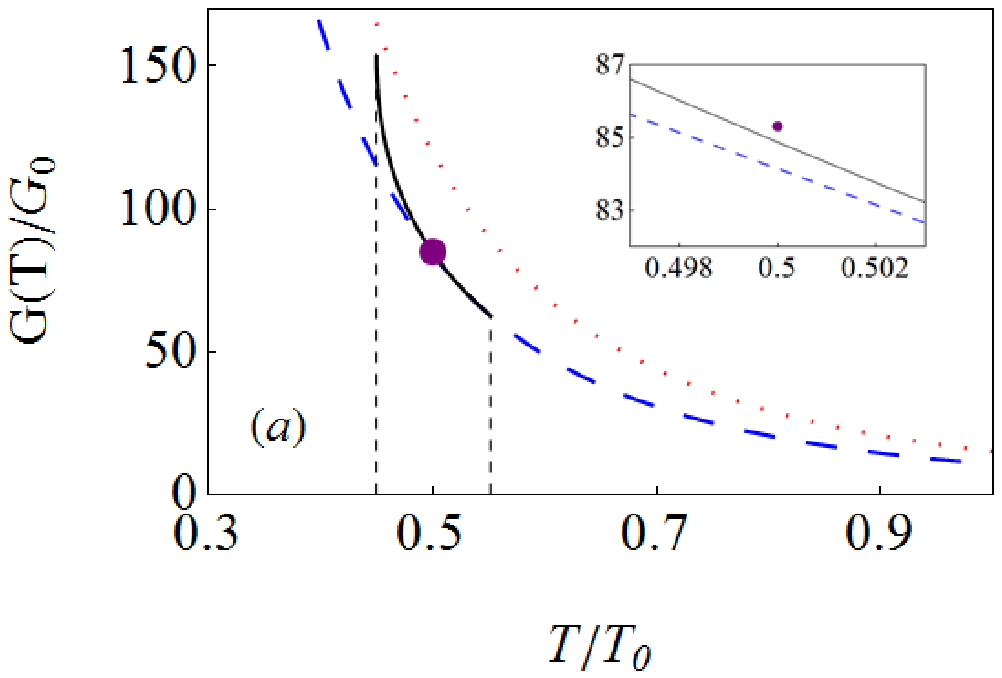}}
\scalebox{0.63}[0.63]{\includegraphics{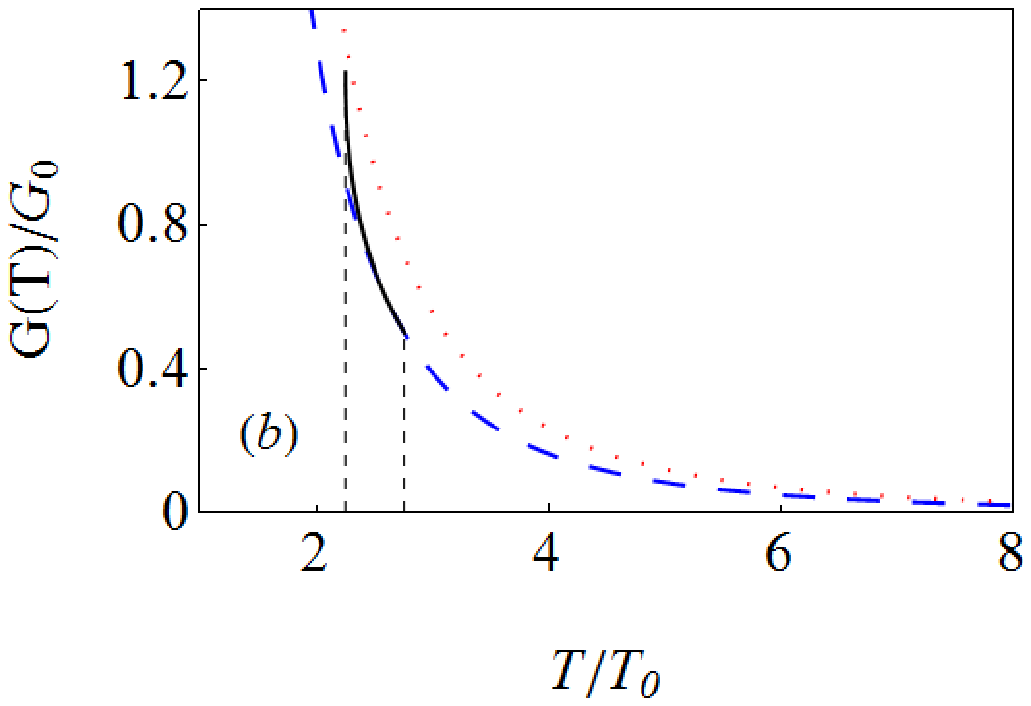}} \caption{(Color
online) G(T) versus final time.
Polynominal ansatz (red dotted line), unbounded optimal (blue dashed
line), bounded optimal (black solid line, the vertical dashed lines
delimitate the time window in Eq. (\ref{timew})), and bang-bang
(purple dot, $T=T_0/2$, see the magnification in the inset). (a)
$\delta=0.5\,d$, (b) $\delta=0.02\,d$, $G_0=\hbar\omega^2\times10^{6}$,
other parameters are the same as in Fig. \ref{figq0}.}
\label{figcom}
\end{center}
\end{figure}

%\subsection{Polynominal Protocol}
%
{\it{Polynomial protocol.}} A simple choice satisfying all boundary
conditions and trap position continuity is a  polynomial ansatz
$q_c(t)=\sum_{n=0}^{5} \beta_n t^n$. The $\beta_n$ can be solved
from the boundary conditions (\ref{conq}) and (\ref{conqdd}) to give
\beq\label{polynomial}
q_c(t)=d(10s^3-15s^4+6s^5),
\eeq
where $s=t/T$, and the corresponding trap trajectory $q_0(t)$ is
obtained from Eq. (\ref{classical}), see Fig. \ref{figq0}.
$G(t)$ becomes
\beq
G(T)=\frac{60md^2}{7T^3}+\frac{(2n+1){\hbar\omega^3}}{4}T,
\eeq
which is depicted in Fig. \ref{figcom} (red dotted line). Short times are
dominated by an inverse-cubic-in-time, frequency-independent term,
and long times by a linear-in-time, $d$-independent  term that
accumulates the effect of noise. A minimum exists at
$T=T_{min}=\sqrt[4]{\frac{720md^2}{7(2n+1)\hbar\omega^3}}$. For the
realistic parameters of the figures, $T_{min}=73.2$ $T_0$, where
$T_0=2\pi/\omega$ is the oscillation period.
This $T_{min}$ is quite a large time (not shown in Fig.
\ref{figcom}) well into the adiabatic regime.\footnote{A general bound for the time-average of the
potential energy $\overline{E_P}$ is \cite{Erik1} $\overline{E_P}\ge 6md^2/(T^4 \omega^2)$.
Thus $\overline{E_P}\approx\hbar\omega$ for the parameters of Fig. 1, requires transport times
$T\ge 36\, T_0$.}
%
%
%
%%%%%%%%%%%%%%%%%%%%%%%%%%%%%%%%%%%%%%%%%%%%%%

%\subsection{Optimal Control}
{\it{Optimal control.}} To minimize $G(T)$ for a given $n$
and fixed transport time $T$, we may apply optimal control theory
with  the cost function
\beq
\label{costfuntion-E} J_{E} = \int_0^T \ddot{q}_c^2(t) dt =
\int^{T}_0 \omega^4 |q_c(t)-q_0(t)|^2 dt,
\eeq
subjected to the conditions (\ref{conq}), and  a constrained
(bounded) displacement $|q_c(t)-q_0(t)| \leq \delta$. This
optimal-control problem was worked out in \cite{transOCT} to
minimize the time-average of the potential energy.
Incidentally, this also minimizes the small effects of fast ion shuttling on the internal states due
to the dc Stark shift \cite{phaseshift} and adverse effects arising from anharmonicities of the trap potentials \cite{2ions}.
The optimal
$q_c(t)$ ($q_0(t)$ follows from Eq.
(\ref{classical}), see Fig. \ref{figq0}) is \cite{transOCT}
\beq q_c (t) = \left\{
\begin{array}{lll}
0,& t \leq 0
\\
\frac{1}{2}\omega^{2} t^2 \delta, & 0<t<t_1
\\
- \frac{1}{6}\omega^2 c_1 (t-\frac{T}{2})^3 +v_0 t+c_2, &
t_1<t<t_1+t_2
\\
d-\frac{1}{2}{\omega^2}(t-T)^2 \delta,& t_1 +t_2 < t < T
\\
d, & t \geq T
\end{array}
\right., \label{qcbc} \eeq
where
%
%\beqa
$c_1= \frac{ 2 \delta}{T - 2t_1}, ~ v_0 = \frac{1}{4} \omega^2
\delta (T +2 t_1),
%\nonumber
%\\
%
%\beqa
%c_2 = \delta T/(T -2 t_1),~
c_2 = \frac{1}{2} (d-v_0 T)$,
%\nonumber
%\\
%\label{time-energy1}
and $t_1 = \frac{T}{2} \left( 1- \sqrt{3} \sqrt{1 - \frac{4
d}{\omega^2 T^2 \delta}}\right)$.
%\eeqa
%
$G(T)$ becomes
\beq
G(T)={\hbar\omega^3}\left[\frac{m\omega}{2\hbar}
\bigg(2\delta^2t_1+\frac{c_1^2t_2^3}{12}\bigg)+\frac{2n+1}{4}T\right].
\label{constr}
\eeq
These equations hold for the time window
\beq\label{timew}
\sqrt{\frac{4d}{\omega^2\delta}}\leq T
\leq\sqrt{\frac{6d}{\omega^2\delta}}.
\eeq
For smaller times there is no solution to the ``bounded control''
optimization problem. For larger times, the solution coincides with
the one for ``unbounded control'', $$q_c=d(3s^2-2s^3).$$ ``Unbounded control''
here means that the displacement is allowed to take any value, and
this ``unbounded'' solution may be applied
to an arbitrarily short time $T$ \cite{Erik1,transOCT,phaseshift}.
The corresponding $G(T)$ is
%%
%\footnote{This solution was also
%Also, Lau and James analyzed in \cite{phase
%shift} the effects of fast ion shuttling on the internal states due
%to the dc Stark shift. They concluded that the main effect is a
%relative phase shift between qubit states. Since the effective
%electric field is proportional to the ion acceleration, the
%magnitude of the shift is proportional to the integral of the
%acceleration over time.
%The corresponding (unbounded control, optimal) $G(t)$ is
%
\beq
G(T)=\frac{6md^2}{T^3}+\frac{(2n+1)\hbar\omega^3}{4}T,
\eeq
%
%%
%\beq \la
%H_0(T)\ra=(n+\frac{1}{2})\hbar\omega+\frac{18md^2}{\omega^2T^4}+\frac{\hbar\omega_0^4}{\omega}\left[\frac{6md^2}{\hbar\omega^3}\frac{1}{T^3}+\frac{2n+1}{4}T\right],
%\eeq
%%
similar in behavior to the polynomial ansatz, see Fig. \ref{figcom}
(blue dashed line). The minimum occurs at
$T=\sqrt[4]{\frac{72md^2}{(2n+1)\hbar\omega^3}}$. For the parameters
of the numerical examples, $T_{min}=66.9$ $T_0$, again well into the
adiabatic regime. The solid lines in Fig. 2 depict $G(T)$ in Eq.
(\ref{constr}) for two values of the constraint.

%%%%%%%%%%%%%%%%%%%%%%%%%%%%%%%%%%%%%%%%%%%%%%%%%%%%%%%%%%%%%%%%%%%%%%%%%%%%%%%%%%
%%%%%%%%%%%%%%%%%%%%%%%%%%%%%%%%%%%%%%%%%%%%%%%%%%%%%%%%%%%%%%%%%%%%%%%%%%%%%%%%%%%%%%%%

%%%%%%%%%%%%%%%%%%%%%%%%%%%%%%%%%%%%%%%%%%%%%%%%%%%%%%%%%%%%%%%%%%%%%%%%%%%%%%%%%%
%%%%%%%%%%%%%%%%%%%%%%%%%%%%%%%%%%%%%%%%%%%%%%%%%%%%%%%%%%%%%%%%%%%%%%%%%%%%%%%%%%%%%%%%

%%%%%%%%%%%%%%%%%%%%%%%%%%%%%%%%%%%%%%%
%\subsection{Bang-bang Protocol }
%
{\it{Bang-bang protocol.}}
Finally let us examine the simple  bang-bang protocol \cite{Alonso}
\beqa q_0(t)=\left\{
  \begin{array}{ll}
    0, & t \leq 0\hbox{}, \\
    d/2, &0<t<T \hbox{}, \\
    d, & t\geq T\hbox{}.
  \end{array}
\right. \eeqa
From Eq. (\ref{classical}), we can solve $q_c(t)$ as
\beqa
\label{qc1}
q_c(t)=\frac{d}{2}-\frac{d}{2}\cos\omega
t+\frac{d(1+\cos\omega T)}{2\sin\omega T}\sin\omega t.
\eeqa
To make $q_c(t)$ satisfy the boundary conditions (\ref{conq}), the
final time must be an odd multiple of a semiperiod, $\omega
T=(2k+1)\pi,~~k=0, 1, 2..$. Now
\beq
G(T)= \frac{m \omega^4 d^2}{16}T+\frac{\hbar\omega^3(2n+1)}{4}T
\eeq
increases linearly with time without a short-time inverse-cubic term
characteristic of the previous protocols. For the minimal time,
$T=T_0/2$, $G(T)$ is just slightly above that for the
unbounded optimal, see the inset in Fig. \ref{figcom} (a). $G(T)$ values for the next
valid times ($3T_0/2$, $5T_0/2$...) are too high and out of scale in
the figure. The unbounded optimal trajectory is quite close to the
bang-bang  one for $T=T_0/2$ but differs significantly from it for
larger times, compare Figs. \ref{figq0} (a) and \ref{figq0} (b).
\subsubsection{Ornstein-Uhlenbeck process}
The Ornstein-Uhlenbeck (OU) noise is
a natural generalization of the Markovian, white noise limit,
with a finite correlation time $\tau$ and a power spectrum of Lorentzian form
%$\sim1/\Omega^{\alpha}$ with $\alpha=2$ and can be written as
%
\beq\label{sou}
S(\Omega)=\frac{D}{2\pi(1+\Omega^2\tau^2)},
\eeq
where $D$ is  the noise intensity.
When $\tau\rightarrow0$, it reduces to  white
noise, and is also instrumental in generating flicker noise (see the following subsection)
by superposing a range of correlation times.
The correlation function corresponding to Eq. (\ref{sou}) is
\beq
\alpha(t)=\frac{D}{2\tau} e^{-t/\tau},
\eeq
so that
\beqa
g_0(t)&=&\frac{D}{2}(1-e^{-
t/\tau}),
\\
g_1(t)&=&\frac{D\tau}{2}\left(1-e^{-t/\tau}-
\frac{t}{\tau}e^{-t/\tau }\right).
\eeqa
The energy in Eq. (\ref{f-energy}) will be
\beq
\label{CFH1}
\nonumber \la H_0(T)\ra_n
=E_n+DG(T),
\eeq
where the excitation energy is $E_e(T)=DG(T)$ and
\beqa
\label{UE}
G(T) =
\nonumber\frac{\hbar\omega^3}{4}\left(2n+1\right)\left(T-\tau+\tau
e^{-T/\tau }\right)
\\\nonumber
+\frac{m}{2}\int_0^T\left[(1-e^{-t/\tau
})\ddot{q}_c^2(t)-\frac{\omega^2 t}{2\tau}e^{-t/\tau
}\dot{q}_c^2(t)\right]dt.
\eeqa
In the small $\tau$ limit, integrating by parts and retaining only linear terms,
\beqa G(T)&=&
\frac{\hbar\omega^3}{2}\left(n+\frac{1}{2}\right)\left(T-\tau\right)
\nonumber\\
&+&\frac{m}{2}\left[\int_0^T\ddot{q}_c^2(t)dt-\tau\ddot{q}_c^2(0)\right].
\eeqa
%
%%
%\beqa \la H_0(T)\ra&=&
%\bigg(n+\frac{1}{2}\bigg)\hbar\omega+\hbar\omega^3\left(n+\frac{1}{2}\right)D\tau^2\left(T-\tau\right)\nonumber\\
%&+&\frac{m}{2}D\tau^2\left[\int_0^T\ddot{q}_c^2(t)dt-\tau\ddot{q}_c(0)\right].
%\eeqa
%%
%
%
%%%%%%%%%%%%%%%%%%%%%%%%%%%%%%%%%%%%%%%%%%%%%%%%%%%%%%%%%%%%%%%%%%%%%%%%%%%%%%%%%%
\begin{figure}[h]
\begin{center}
\scalebox{0.72}[0.72]{\includegraphics{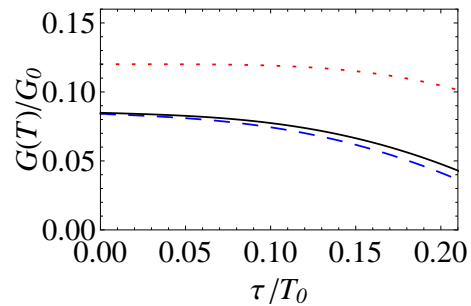}} \caption{(Color
online) G(T) for Ornstein-Uhlenbeck noise versus correlation time. Polynomial ansatz (red dotted
line), unbounded optimal (blue dashed line), and bounded optimal (black
solid line).
%, and bang-bang (purple dash-point).
$\delta=0.005\,d$, $T=5T_0$, $G_0=\hbar \omega^2\times10^{6}$ and
other parameters are the same as in Fig. \ref{figq0}.}
\label{fig3UE}
\end{center}
\end{figure}
%%%%%%%%%%%%%%%%%%%%%%%%%%%%%%%%%%%%%%%%%%%%%%%%%%%%%%%%%%%%%%%%%%%%%%%%%%%%%%%%%%%
The two  correcting terms proportional to
$\tau$ are negative so that the noise effect is reduced with respect to white noise.
In Fig. \ref{fig3UE} we plot $G(T)$ versus
correlation time using the polynomial protocol and the protocols optimized for white noise.
%%%%%%%%%%%%%%%%%%%%%%%%%%%%%%%%%%%%%%%%%%%%%%%%%%%%%%%%%%%%%%%%%%%%%%%%%%%%%%%%%%%%%%%%%%%
%
%
%
%
\subsubsection{Flicker noise}
Flicker noise, with $\sim 1/\Omega$ spectrum in a range $\Omega_2<\Omega<\Omega_1$,
may be modeled by summing over Lorentzian (Ohrnstein-Uhlenbeck)
noises \cite{Hooge,watanabe} with proper statistical weights.
Specifically we consider \cite{Hooge}
\beq
\label{alfli}
\alpha(t)=\frac{C}{\ln(\tau_2/\tau_1)}\int_{\tau_1}^{\tau_2}\frac{1}{\tau}e^{-t/\tau}d\tau,
\eeq
where
%$\tau_1=1/(\Omega_1)$ and $\tau_2=1/(\Omega_2)$, and
$C={\cal{E}}[
x^2(t)]=\alpha(0)$. Using Eq. (\ref{som}), the corresponding power spectrum takes the form
\beqa
S(\Omega)&=&\frac{C}{\pi \ln(\tau_2/\tau_1)}\int_{\tau_1}^{\tau_2} \frac{d\tau}{1+\Omega^2\tau^2}
\nonumber\\
&=&\left\{
        \begin{array}{ll}
          \frac{C(\tau_2-\tau_1)}{\pi\ln(\tau_2/\tau_1)}, & \hbox{} \Omega\ll\Omega_2,\\
          \frac{C}{2\ln(\tau_2/\tau_1)}\frac{1}{\Omega}, & \hbox{} \Omega_2\ll\Omega\ll\Omega_1, \\
          \frac{C(\tau_2-\tau_1)}{\pi\ln(\tau_2/\tau_1)}\frac{1}{\tau_1\tau_2\Omega^2}, &
\hbox{}\Omega\gg\Omega_1.
        \end{array}
      \right.
\eeqa
where $\Omega_{1,2}=(2\pi)/\tau_{1,2}$.
The spectrum is white if the frequency is below $\Omega_2$ and decays as $1/\Omega^2$
above $\Omega_1$.
Eq. (\ref{alfli}) leads to
\beqa
g_0(t)&=&\frac{C}{\ln(\tau_2/\tau_1)}\left[\tau-\tau
e^{-t/\tau}-t
Ei\left(\frac{-t}{\tau}\right)\right]_{\tau_1}^{\tau_2},
\\
g_1(t)&=&\frac{C}{2\ln(\tau_2/\tau_1)}
\nonumber\\
&\times&\!\left[\tau^2(1\!-\!e^{t/\tau})
\!-\!t\tau
e^{-t/\tau}\!-\!t^2 Ei\left(\frac{-t}{\tau}\right)\right]_{\tau_1}^{\tau_2}\!\!.
\eeqa
Here $Ei[-x]=\int_{-\infty}^{-x}(e^{t}/t) dt$ with $x>0$, which behaves as
$Ei[-x]\simeq-e^{-x}/x$ for $x\rightarrow\infty$, and $Ei[-x]\simeq
\gamma_E +\ln x$ for $x\rightarrow0$, where $\gamma_E$ is Euler's constant.
The energy (\ref{f-energy}) takes the form
\beq
 \la H_0(T)\ra_n=E_n+\frac{2C(\tau_2-\tau_1)}{\ln(\tau_2/\tau_1)}G(T),
\eeq
where the excitation energy is
$E_e(T)=\frac{2C(\tau_2-\tau_1)}{\ln(\tau_2/\tau_1)}G(T)$ and
\begin{widetext}
\beqa\label{flicker energy}G(T)=
\frac{\hbar\omega^3}{4(\tau_2-\tau_1)}\left(n+\frac{1}{2}\right)
\left[\tau
T(2-e^{-T/\tau})+\tau^2(e^{-T/\tau}-1)-T^2Ei\bigg(\!\!\!-\frac{T}{\tau}\bigg)\right]_{\tau_1}^{\tau_2}
\nonumber
%H_0(T)\ra=(n+\frac{1}{2})\hbar\omega+\frac{\hbar\omega^3}{\ln(\tau_2/\tau_1)}(n+\frac{1}{2})\int_{\tau_1}^{\tau_2}
%(T-\tau+\tau e^{-T/\tau})d\tau\nonumber
\\
+\frac{m
}{2(\tau_2-\tau_1)}\int_0^T\left\{\ddot{q}_c^2(t)\left[\tau-\tau
e^{-t/\tau }-t
Ei\bigg(\!\!\!-\frac{t}{\tau}\bigg)\right]_{\tau_1}^{\tau_2}+\frac{\omega^2
t}{2}\dot{q}_c^2(t)Ei\bigg(\!\!\!-\frac{t}{\tau}\bigg)\bigg|_{\tau_1}^{\tau_2}\right\}
dt. \eeqa
\end{widetext}

\begin{figure}[h]
\begin{center}
\scalebox{0.72}[0.72]{\includegraphics{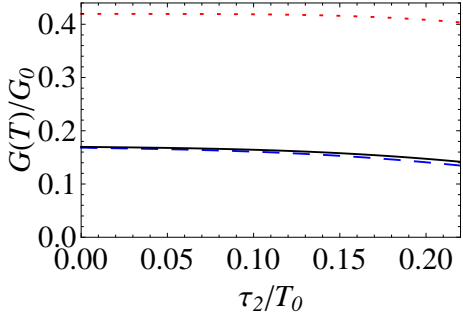}}
%\scalebox{0.47}[0.47]{\includegraphics{fig4FE-b.eps}}
\caption{(Color online) $G(T)$ for flicker noise
versus upper limit of the interval of correlation times $\tau_2$.
Polynomial ansatz
(red dotted line), unbounded optimal (blue dashed line), and bounded
optimal (black solid line).
%, and bang-bang (purple dashed-point).
$\delta=0.5\,d$,
%(b) $\delta=0.005\,d$,
$G_0=\hbar\omega^2 \times10^{6}$, $\tau_1=1\times10^{-10}$ $s$,
$T=5T_0$, and other parameters are the same as in Fig. \ref{figq0}.}
\label{figFE}
\end{center}
\end{figure}

For  $\tau_2/T\ll1$ and $\dot{q_c}(0)=0$, we find, integrating by parts,  the approximation
\beqa \label{f-f energy}
G(T)&\simeq&
\frac{\hbar\omega^3}{2}\left(n+\frac{1}{2}\right)\left(T-\frac{\tau_2+\tau_1}{2}\right)
%+\frac{\tau_2^2
%e^{-T/\tau_2}-\tau_1^2 e^{-T/\tau_1}}{2(\tau_2-\tau_1)}\right)
\nonumber
\\
&+&\frac{m}{2}\left[\int_0^T\ddot{q}_c^2(t)
dt-\frac{\tau_2+\tau_1}{2}\ddot{q}_c^2(0)\right],
\eeqa
with a small correction to the white noise case similar to the one found for
Ornstein-Uhlenbeck noise.
Fig. \ref{figFE} depicts $G(T)$ versus $\tau_2$ for the polynomial protocol and the protocols
optimized in the Markovian limit.
%%
%%%%%%%%%%%%%%%%%%%%%%%%%%%%%%%%%%%%%%%%%%%%%%%%%%%%%%%%%%%%%%%%%%%%%%%%%%%%%%%%%%
%%%%%%%%%%%%%%%%%%%%%%%%%%%%%%%%%%%%%%%%%%%%%%%%%%%%%%%%%%%%%%%%%%%%%%%%%%%%%%%%%%%%%%%%
%
%
%
%
\subsection{Position noise}
In this subsection we define $L=K(\hat{q}-q_0)$
in Eqs. (\ref{hami}) and (\ref{ME}) to simulate
the effect of the environment on a fluctuating trap position.
The master equation (\ref{ME}) takes the form
%%
%\beq \bar{O}(t)=g_0(t)L-\frac{1}{m}g_1(t)\hat{p}.\eeq
%%
%
\beqa
\frac{d}{dt}\rho=&-&\frac{i}{\hbar}[H_0,\rho]-\frac{K^2}{\hbar^2}g_0(t)\left[\hat{q}-q_0,[\hat{q}-q_0,\rho]\right]
\nonumber\\
&+&\frac{K^2}{m\hbar^2}g_1(t)\left[\hat{q}-q_0,[\hat{p},\rho]\right].
\label{mepos}
\eeqa
Using the same time-dependent perturbation theory approach as in the
previous section, the density matrix is
\beqa
\rho(T)&\simeq&\rho_0(T)+\frac{K^2}{\hbar^2}\int_0^Tg_0(t)\widetilde{U}_0(T,t)\widetilde{J}_2(t)\rho_0(t)dt\nonumber
\\\nonumber
&+&\frac{K^2}{m\hbar^2}\int_0^Tg_1(t)\widetilde{U}_0(T,t)  [\hat{q},[\hat{p},\rho_0(t)]] dt,
\eeqa
where
%$\widetilde{J}_5(t)\rho_0(t)=$,
the system energy is
\beqa
\label{CPH}\nonumber
\la H_0(T)\ra_n &=& tr[H_0(T)\rho(T)]
\simeq \la\Psi_n(T)|H_0(T)|\Psi_n(T)\ra
\\
\nonumber
&+&\frac{K^2}{\hbar^2}\!\!\int_0^T\!\!\!g_0(t)\la\Psi_n(t)|\widetilde{J}_2(t)H'(t)|\Psi_n(t)\ra
dt
\\\nonumber
&+&\frac{K^2}{m\hbar^2}\!\!\int_0^T\!\!\!g_1(t)\la\Psi_n(t)|[\hat{p},[\hat{q},H'(t)]]|\Psi_n(t)\ra
dt
\\
&=&E_n+\frac{K^2}{m}\int_0^Tg_0(t)dt.
\eeqa
The excitation energy at the final
time is independent of the trap trajectory, and depends only on the
transport time. The only strategy left to minimize the effect of
position fluctuations is to speed up the transport making $T$ as
small as possible. The independence on the trajectory may  be
understood already at classical level from the solution of Eq.
(\ref{classical}), $q_c(t)=q_0(t)-\int_0^t dt'
\dot{q}_0(t')\cos[\omega(t-t')]$. Note that a deviation from
$q_c(t)$ due to a modified trajectory $q_0+\delta q_0$ depends only
on $\delta q_0$ and its time derivative, not on $q_0$ itself.
As a consequence, studies of excitation or heating rates for non-shuttling traps
are directly applicable \cite{Savard1,Savard2,Milburn,Lamoreaux}.
%%%%%%%%%%%%%%%%%%%%%%%%%%%%%%%%%%%%%%%%%%%%%%%%%%%%%%%%%%%%%
%
%
%
\section{Systematic spring constant error} \label{error}
Assume that the trap trajectory is designed for a given spring constant
$\omega^2$, but the actual one is different, $\omega^2(1+\lambda)$.
$\lambda$ may change from run to run
but remain constant throughout the transport time.
This is quite common as a consequence of experimental drifts and imperfect calibration.
In current experiments it is likely to dominate other imperfections.
Our objective here is to
determine the induced excitation and  to find trap trajectories that minimize the excitation
in a range of $\lambda$ around 0.
The system Hamiltonian is
\beq
H (t)=\frac{\hat{p}^2}{2 m} + \frac{1}{2}m \omega^2(1+\lambda)
[\hat{q}-q_0(t)]^2,
\eeq
where $\lambda$ is the relative error in the spring constant. For
the actual frequency, the auxiliary equation is
\beq
\label{newclassical}\ddot{Q}_{c1}(t)+\omega_1^2(Q_{c1}-q_0)=0,
\eeq
with $\omega_1^2=\omega^2(1+\lambda)$. We define
$Q_{c1}(t)=q_c(t)+f(t)$. Combining Eqs. (\ref{classical}) and
(\ref{newclassical}),  $f(t)$ satisfies
\beq
\ddot{f}(t)+\omega^{2}_1f(t)=\lambda\ddot{q}_c(t),
\eeq
which is solved by
\beqa
f(t)&=&\frac{\lambda}{\omega_1}\sin(\omega_1t)\int_0^t\ddot{q}_c(t')\cos(\omega_1t')dt'\nonumber
\\&-&\frac{\lambda}{\omega_1}\cos(\omega_1t)\int_0^t\ddot{q}_c(t')\sin(\omega_1t')dt'.
\eeqa
For the new frequency $\omega_1$ and trajectory $Q_{c1}$, the exact
energy of the system takes the form
\beqa
\label{FHerror}\la H(T)
\ra_n&=&\left(n+\frac{1}{2}\right)\hbar\omega_1+E_e(T),
\eeqa
where $E_e$ is the excitation energy,
\beqa
\label{FHerror}
E_e(T)&=&\frac{m\lambda^2}{2}\left[\int_0^T\ddot{q}_c(t)\cos(\omega_1t)dt\right]^2
\nonumber
\\
&+&\frac{m\lambda^2}{2}\left[\int_0^T\ddot{q}_c(t)\sin(\omega_1t)dt\right]^2.
\eeqa
To suppress the excitation energy, the trajectory $q_c(t)$
has to satisfy the conditions
\beq
\label{qcsin} \int_0^T\ddot{q}_c(t)\cos(\omega_1t)dt=0,~~~
\int_0^T\ddot{q}_c(t)\sin(\omega_1t)dt=0.
\eeq
%
%%%%%%%%%%%%%%%%%%%%%%%%%%%%%%%%%%%%%%%%%%%%%%%%%%%%%%%%%%%%%%%%%%%%%%%%%%%%%%%%%%%
\begin{figure}[h]
\begin{center}
\scalebox{0.7}[0.7]{\includegraphics{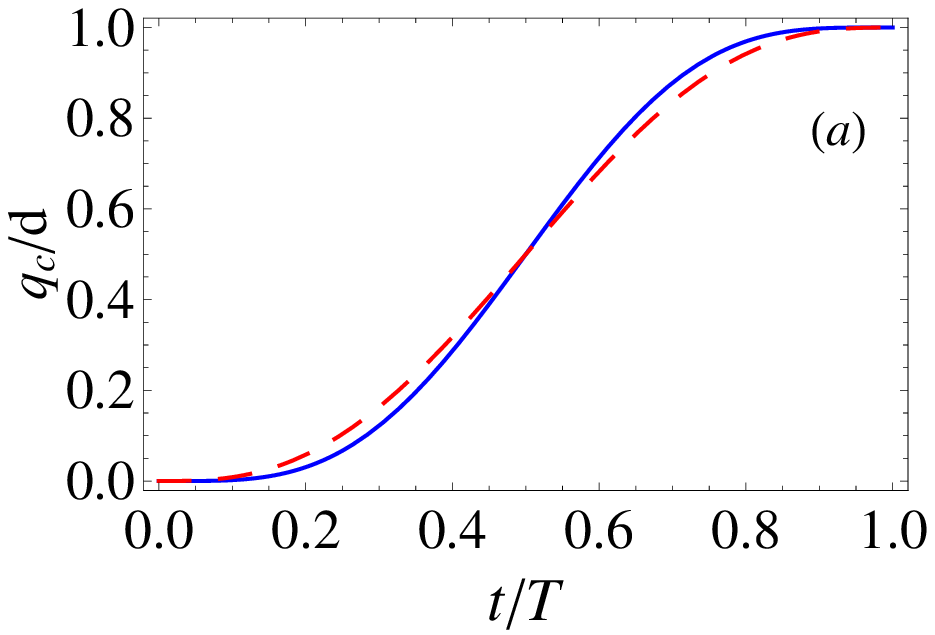}}
\scalebox{0.7}[0.7]{\includegraphics{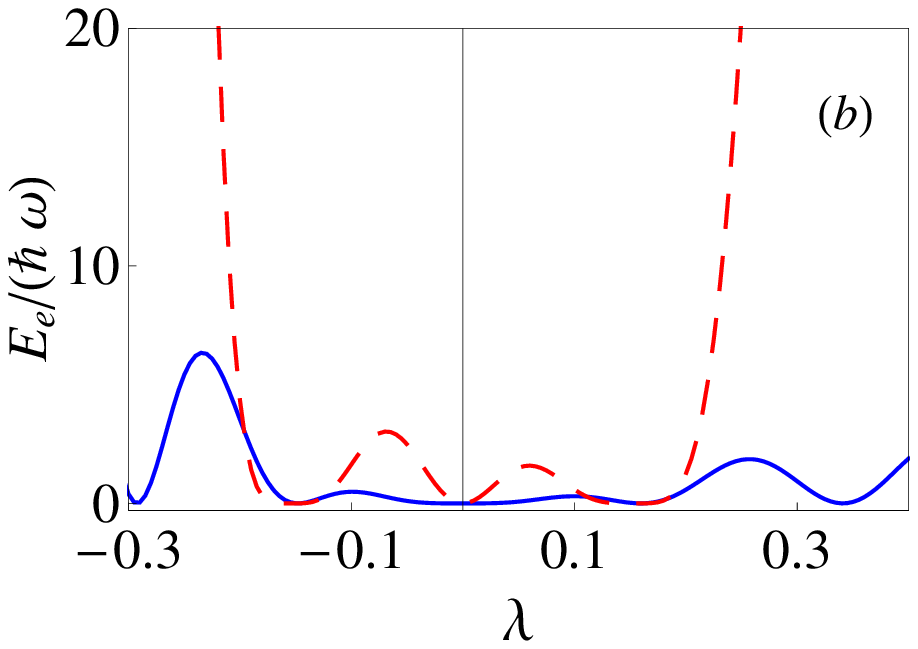}}
\caption{(Color online)  (a) $q_c(t)$ versus $t$; (b) excitation energy versus
$\lambda$. Dashed red line:
quintic polynomial (\ref{polynomial}); solid blue line: seventh order polynomial in Eq. (\ref{qc8}).
$T=6.5\,T_0$ and other parameters are the same as in Fig.
\ref{figq0}.} \label{fig3}
\end{center}
\end{figure}
%%%%%%%%%%%%%%%%%%%%%%%%%%%%%%%%%%%%%%%%%%%%%%%%%%%%%%%%%%%%%%%%%%%%%%%%%%%%%%%%%%%%%%%%%
%
We approximate $\cos(\omega_1 t)\simeq\cos(\omega t)$ and
$\sin(\omega_1t)\simeq\sin(\omega t)$ to keep only quadratic terms in $\lambda$
in Eq. (\ref{FHerror}),
and assume for $q_c$ a seventh order polynomial
\beq
\label{qc8}
q_c(t)=\sum_{n=0}^{7}c_nt^n
\eeq
to satisfy the six conditions
in Eqs. (\ref{conq}), (\ref{conqdd}) and
\beq
\label{con-error}\int_0^T\ddot{q}_c(t)\cos(\omega t)\,dt=0,~~~
\int_0^T\ddot{q}_c(t)\sin(\omega t)\,dt=0.
\eeq
Doing the integrals formally, in terms of the unknown coefficients,
we end up with a system of eight equations with eight unknowns (the $c_n$),
which can be solved, but the expressions for the $c_n$ are too lengthy to be
displayed here.
%
%where $s=t/T$.
The corresponding  $q_0$ is obtained from Eq. (\ref{classical}).
In Fig. \ref{fig3} we have plotted the seventh order $q_c$
in Eq. (\ref{qc8}) and the simplest
quintic polynomial ansatz (\ref{polynomial}), as well as the corresponding excitation energies.
The protocol based on Eq. (\ref{qc8}) is more robust, i.e., it leads to smaller excitations when the actual trap frequency
does not have the expected value.
Alternative robustification schemes are possible adapted to specific needs, for example,
imposing zero or minimal excitation at a discrete number of values of $\lambda$ in a given interval, see e.g.
\cite{oldPRL} for a similar approach applied to maximize the absorption of complex potentials.

Time-scaling errors are shown to be equivalent to
spring-constant systematic errors in Appendix B, so the same strategies used here
may be used in that case.
\section{Discussion}
%
%
%
%
%
% {\it{Discussion}---}
%
%
In this paper we have examined the excitation energy due to
spring-constant noise/error and position noise in ions transported by a
moving harmonic trap. We consider families of trajectories without
final excitation in the noiseless limit and select optimal trap
trajectories that minimize heating when the noise applies. For fixed
shuttling time $T$, this selection is only possible for
spring-constant noise/error, since for position noise the final energy
increases linearly with $T$ but does not depend on any other feature
of the trap trajectory.

We find an additional beneficial feature of the trajectories
that minimize the effect of spring-constant noise even in the case that position
noise is dominant:
These trajectories minimize the time-average of the potential energy
\cite{Erik1,transOCT}, thus adverse effects of
anharmonicity \cite{Mainz1, 2ions} are suppressed.

Apart from trap trajectories with sudden, finite position jumps (optimal trap trajectories unconstrained or constrained by a maximum ion
displacement with respect to the center
of the trap, and simple bang-bang trajectories)  we have as well considered smooth polynomial trajectories.
For very short shuttling times (half an oscillation  period) optimal control and bang-bang
solutions display a reduced noise sensitivity, although they imply the technical challenge of
implementing sudden trap jumps. At moderate times and beyond (five oscillations or more) the bang-bang approach
produces too much excitation and the polynomial behaves similarly to the optimal trajectory.

Advances in the fabrication of micro structured ion traps and fast control electronics
have allowed to experimentally reach the limits of adiabacity, thus the proposed protocols may be tested and the respective noise-sensitivity verified.
Envisaged experiments at shuttle times of the order of an oscillation period \cite{Alonso}
require changes of the trapping potential on timescales much shorter than the period corresponding to the
trap frequency. At such fast temporal changes of the control voltages, the cut-off frequency for
noise filtering elements must be very high, and thus we expect that it might be increasingly difficult to reach a low noise level.
As additionally the noise sensitivity of the shuttling results is increasing at fast timescales, the importance of noise-suppression by trajectory design becomes obvious.
In the well-controlled setting of an ion trap, one may experimentally investigate the schemes with artificial injected
designed noise \cite{arti1,arti2}.
It is in experimental reach to design the spectral properties of a noise source and verify the predicted effects.
The accuracy of sideband spectroscopy to determine the excess energy  has reached sub-phonon level, such that even
small optimization effects would be visible.

{\it{Acknowledgments}---}
This work was supported by the
Grants No. 61176118, 12QH1400800, 13PJ1403000, 2013310811003,
IT472-10, FIS2009-12773-C02-01, UFI 11/55, and the Program for Professor of Special Appointment
(Eastern Scholar) at Shanghai Institutions of Higher Learning.
This research was also funded by the Office of the Director of National Intelligence (ODNI),
Intelligence Advanced Research Projects Activity (IARPA), through the Army Research Office grant W911NF-10-1-0284. All statements of fact, opinion or conclusions contained herein are those of the authors and should not be construed as representing the official views or policies of IARPA, the ODNI,
or the US Government.
%National Natural Science Foundation
%of China (Grant No. 61176118), the Shanghai Rising-Star and Pujiang
%Program (Grant Nos. 12QH1400800 and 13PJ1403000), the Specialized
%Research Fund for the Doctoral Program of Higher Education (Grant
%No. 2013310811003), the Program for Professor of Special Appointment
%(Eastern Scholar) at Shanghai Institutions of Higher Learning, the
%Basque Government (Grant No. IT472-10), Ministerio de Econom\'\i a y
%Competitividad (Grant No. FIS2009-12773-C02-01), the UPV/EHU under
%program UFI 11/55.
%, and COST programme, under grant number
%COST-C12.0118.
%
%
%
%
\appendix
\section{Closed equations for the moments}
The quadratic and linear operators involving position and momentum
form a dynamical Lie algebra (the Hamiltonian is a member of this
algebra) for the Hamiltonians that describe spring constant noise
and position noise. This leads to closed equations for the
corresponding moments \cite{Milburn}, which is interesting numerically, as the results are not perturbative in noise intensity.
Also, physical consequences follow
without even solving the system as we shall see.

For spring constant noise, the expectation values of
position and momentum operators and their quadratic combinations
satisfy, using Eq. (\ref{f-masterEq}),
\begin{widetext}  \beq \label{f-closeEq}\frac{d}{dt}\!\left(\!\!
                    \begin{array}{c}
                      \la \hat{q}^2\ra \\
                      \la \hat{p}^2\ra \\
                      \la \hat{q}\hat{p}+\hat{p}\hat{q}\ra \\
                      \la \hat{q}\ra \\
                      \la \hat{p}\ra \\
                    \end{array}
                  \!\!\right)=M_{S}\left(\!\!
                    \begin{array}{c}
                    \la \hat{q}^2\ra \\
                      \la \hat{p}^2\ra \\
                      \la \hat{q}\hat{p}+\hat{p}\hat{q}\ra \\
                      \la \hat{q}\ra \\
                      \la \hat{p}\ra \\
                    \end{array}
                  \!\!\right)+\left(\!\!
                    \begin{array}{c}
                      0 \\
                      8\hbar^2q_0^2g_0(t)\\
                      \frac{8\hbar^2}{m}q_0^2g_1(t) \\
                      0 \\
                     m\omega^2 q_0-\frac{4\hbar^2}{m}q_0g_1(t) \\
                    \end{array}
                  \!\!\right),
\eeq
where
\beqa \label{f-matrix}
 M_{S}= \left(
               \begin{array}{ccccc}
               0 & 0 & \frac{1}{m} & 0 & 0 \\
               8\hbar^2g_0(t) & 0 & -m\omega^2 & -16\hbar^2q_0g_0(t)& 2m\omega^2 q_0 \\
               -2m\omega^2+\frac{16\hbar^2}{m}g_1(t) & \frac{2}{m} & 0 & 2m\omega^2 q_0-\frac{8\hbar^2}{m}q_0g_1(t) & 0 \\
               0 & 0 & 0 & 0 & \frac{1}{m}\\
               0 & 0 & 0 & -m\omega^2+\frac{4\hbar^2}{m}g_1(t) & 0 \\
               \end{array}
               \right).\nonumber\\
\eeqa \end{widetext}
%
%\section{position noise: matrix}
For position noise, Eq. (\ref{mepos}), the expectation values satisfy
%
%\begin{widetext}
\beqa
\label{pequation}
\frac{d}{dt}\!\!\left(\!\!\!
                    \begin{array}{c}
                      \la \hat{q}^2\ra \\
                      \la \hat{p}^2\ra \\
                      \la \hat{q}\hat{p}+\hat{p}\hat{q}\ra \\
                      \la \hat{q}\ra \\
                      \la \hat{p}\ra \\
                    \end{array}\!\!\!
                  \right)\!=\!M_{P}\!\left(\!\!\!
                    \begin{array}{c}
                     \la \hat{q}^2\ra \\
                      \la \hat{p}^2\ra \\
                      \la \hat{q}\hat{p}+\hat{p}\hat{q}\ra \\
                      \la \hat{q}\ra \\
                      \la \hat{p}\ra \\
                    \end{array}\!\!\!
                  \right)\!\!+\!\!\left(\!\!\!
                    \begin{array}{c}
                     0 \\
                      2K^2g_0(t) \\
                      2K^2g_1(t)/m\\
                      0 \\
                      m\omega^2 q_0 \\
                    \end{array}\!\!\!
                  \right)
\eeqa
%\rm{where}\;\;\;
where
\beq
 M_{P}= \left(
               \begin{array}{ccccc}
               0 & 0 & 1/m & 0 & 0\\
               0 & 0 & -m\omega^2 & 0 & 2m\omega^2 q_0  \\
               -2m\omega^2 & 2/m & 0 & 2m\omega^2 q_0 & 0 \\
               0 & 0 & 0 & 0 & 1/m \\
               0 & 0 & 0 & -m\omega^2 & 0 \\
               \end{array}
               \right).
\eeq
%\nonumber\\\eeqa \end{widetext}
%
For colored or white position noise, the average position and momenta are
not affected by the noise.
\section{Time scaling}
We analyze here a systematic error in the clock used to design the
trap trajectory
so that instead of $q_0(t)$, the  implemented trajectory is $q_0(\varepsilon
t)$.
The Hamiltonian is
\beq
H(t)=\frac{\hat{p}^2}{2m}+\frac{1}{2}m\omega^2[\hat{q}-q_0(\varepsilon
t)],
\eeq
and the Schr\"{o}dinger equation $i\hbar\partial\Psi(t)/\partial
t=H(t)\Psi(t)$ can be rewritten as
\beq \label{schordinger} i\hbar\frac{\partial\Phi(\tau)}{\partial
\tau}=H'(\tau)\Phi(\tau), \eeq
where $\tau=\epsilon t$, $\Phi(\tau)=\Psi(t)$, and
\beq
H'(\tau)=\frac{p^2}{2
m'}+\frac{1}{2}m'\omega'^2(\hat{q}-q_0(\tau)),
\eeq
with $m'=\varepsilon m$, and $\omega'=\omega/\varepsilon$.
Since $q_0(\tau)$ is designed for $\omega$, time scaling errors reduce formally to systematic spring-constant errors,
and their effect can be suppressed or mitigated in the same manner.

\end{document}